\documentclass[aps,prd,nofootinbib,showkeys,floatfix,twocolumn]{revtex4-1}
\usepackage{amsmath}
\usepackage{amssymb}
\usepackage{graphicx}
\usepackage{xspace}
\usepackage{bbm} 

\usepackage[utf8]{inputenc}

\newcommand{\eVdist}{\kern-0.06em}

\newcommand{\Tev}{\text{Te\eVdist V}}

\newcommand\SARAH{{\tt SARAH}\xspace}
\newcommand\SPheno{{\tt SPheno}\xspace}

\newcommand\Mathematica{{\tt Mathematica}\xspace}

\newcommand{\be}{\begin{equation}}
\newcommand{\ee}{\end{equation}}
\newcommand{\bea}{\begin{eqnarray}}
\newcommand{\eea}{\end{eqnarray}}

\begin{document}
\title{On the MSSM Higgsino mass and fine tuning}

\hfill \parbox{5cm}{\flushright DESY-16-062\\ OUTP-16-06P \\ CERN-TH-2016-072}

\newcommand{\AddrCERN}{%
Theoretical Physics Department, CERN, Geneva, Switzerland
}
\newcommand{\AddrOxford}{%
Rudolf Peierls Centre for Theoretical Physics, University of Oxford,\\
 1 Keble Road, Oxford OX1 3NP, UK}
\newcommand{\AddrDESY}{%
Deutsches Elektronen-Synchrotron DESY, \\
  Notkestra\ss e 85, D-22607 Hamburg, Germany}

 \author{Graham G.~Ross}
 \email{g.ross1@physics.ox.ac.uk}
 \affiliation{\AddrOxford}

 \author{Kai Schmidt-Hoberg}
 \email{kai.schmidt-hoberg@desy.de}
 \affiliation{\AddrDESY}

 \author{Florian Staub}
 \email{florian.staub@cern.ch}
 \affiliation{\AddrCERN}

\begin{abstract}
It is often argued that low fine tuning in the MSSM necessarily requires a rather light Higgsino. 
In this note we show that this need not be the case when a more complete set of soft SUSY breaking mass terms are included. In particular an Higgsino mass term, that correlates the $\mu-$term contribution with the soft SUSY-breaking  Higgsino masses, significantly reduces the fine tuning even for Higgsinos in the $\Tev$ mass range where its relic abundance means it can make up all the dark matter. 
\end{abstract}

\maketitle

\section{Introduction}

Our expectation of what to find beyond the Standard Model (SM) of particle physics has largely been shaped by naturalness arguments,
and arguably low energy supersymmetry emerged as the prime candidate for BSM physics.
Fine tuning considerations give us a handle to judge (i) which classes of models are (more) natural but also (ii) for a given model which parameter choices
are preferred.
It has been realised long ago that the $\mu$ term plays a
special role in fine tuning considerations. The RGE evolution is very mild and the fine tuning with respect to $\mu$ can be estimated as
\begin{equation}
\Delta_\mu \sim  \frac{2\mu^2}{M_Z^2}
\end{equation}
which implies that for a natural theory with fine tuning $\Delta_\mu < 100$ the value of $\mu$ should not exceed a few hundred GeV.
As the Higgsino mass in the usual MSSM is roughly given by $\mu$, this has led to the belief that a natural theory necessarily requires a rather light Higgsino,
see e.g.~\cite{Baer:2012up,Baer:2012uy,Baer:2012cf,CahillRowley:2012rv,Feng:2012jfa,Kang:2012sy,Baer:2013gva,Baer:2013vpa,Kowalska:2013ica,Baer:2014ica}. 
It is of crucial importance to evaluate whether this conclusion is true, as light Higgsinos start to be established as one of the main tell tale signs for naturalness 
within the community and search strategies are developed accordingly.

A light Higgsino typically means that it is the lightest supersymmetric particle (LSP).
As the annihilation cross section of Higgsinos is sizeable, dark matter is typically underproduced in this case.
While this is experimentally viable, because dark matter could consist of several components (there might e.g.\ be an additional component of axion dark matter), 
it would be nice to saturate the relic abundance with Higgsinos only. The Higgsino relic abundance depends on the Higgsino mass $m_{\tilde H}$
and the correct relic abundance is achieved for $m_{\tilde H} \sim 1$~TeV. This however seems to be in strong tension with naturalness arguments.

In this note we show that in the MSSM Higgsino masses of about 1~TeV can be achieved with low fine tuning.
The key insight is that in addition to the usual $\mu$ term a SUSY breaking Higgsino mass term can be present.
As discussed below although such a term can readily be generated nevertheless it is almost always discarded. The reason is that it can 
be reabsorbed into other parameters of the model and hence seems superfluous. While this is true with respect to the particle spectrum,
the inferred values for the fine tuning can differ significantly. Accordingly, the conclusions with regards to the particle spectrum based
on fine tuning considerations change as well; in particular a TeV scale Higgsino might well be natural.

This letter is organises as follows. In Section~\ref{sec:model} we introduce the new ingredients we consider in the context of the MSSM and discuss in detail the structure and possible origin of the Higgsino mass term. In Section~\ref{sec:FT} we give an approximate relation between the fine-tuning in the model 
and the new soft-terms before we perform, in Section~\ref{sec:num}, a purely numerical study of the fine-tuning. We conclude in 
Section~\ref{sec:conclusion}.

\section{The MSSM with the full set of soft terms}
\label{sec:model}
Let us consider the MSSM extended by the following non-holomorphic soft-terms,  ``soft" in the sense that they do not lead to quadratic divergences at radiative order \cite{Girardello:1981wz},
\begin{align}
\mathcal{L}_{NH} = & T'_{u,ij}  H_d^* \tilde u_{R,i}^* \tilde q_j +T'_{d,ij}  H_u^* \tilde d_{R,i}^* \tilde q_j + \nonumber \\
\label{eq:LNH}
 & T'_{e,ij}  H_u^* \tilde e_{R,i}^* \tilde l_j  + \mu' \tilde H_d \tilde H_u \; +  \text{h.c.}
\end{align}
The potential origin of these terms can either be spontaneous
SUSY breaking within gravity mediation \cite{Martin:1999hc}, strongly
coupled SUSY gauge theories \cite{ArkaniHamed:1998wc}, or they are radiatively-generated
in $N=2$ and $N=4$ SUSY gauge theories \cite{deWit:1996kc,Bellisai:1997ck,Dine:1997nq,GonzalezRey:1998gz,Buchbinder:1998qd}. In \ref{sec:Higgsinomass} we give an explicit example for the case of the Higgsino mass term.

While $\mu'$ enters the neutralino mass matrix
\begin{equation} 
m_{\tilde{\chi}^0} = \left( 
\begin{array}{cccc}
M_1 &0 &-\frac{1}{2} g_1 v_d  &\frac{1}{2} g_1 v_u \\ 
0 &M_2 &\frac{1}{2} g_2 v_d  &-\frac{1}{2} g_2 v_u \\ 
-\frac{1}{2} g_1 v_d  &\frac{1}{2} g_2 v_d  &0 &- {\mu'}  - \mu \\ 
\frac{1}{2} g_1 v_u  &-\frac{1}{2} g_2 v_u  &- {\mu'}  - \mu  &0\end{array} 
\right) 
 \end{equation} 
the mass matrices for all scalars as well as the two minimisation conditions $\frac{\partial V}{\partial v_i} =0$ ($i=u,d$) are 
not changed compared to the MSSM without the terms given in eq.~(\ref{eq:LNH}). 
Thus, the dependence of $M^2_{Z}$ on the SUSY parameters and $\tan\beta=\frac{v_u}{v_d}=t_\beta$ is as usual
\begin{align}
\frac{M^2_Z}{2} =&  \frac{\mu ^2+ m_{H_d}^2+t_\beta^2 \left(-\left(\mu ^2+m_{H_u}^2\right)\right)}{t_\beta^2-1} \nonumber \\
 & \quad \simeq -\mu(Q)^2 - m_{H_u}(Q)^2
\end{align}
where in the last step we explicitly show the dependence on the scale $Q$ at which the parameters are determined. In the following, unless otherwise stated, we take this to be the SUSY breaking scale. 
Some phenomenological consequences of the additional soft-terms were analysed in Refs.~\cite{Demir:2005ti,Un:2014afa}.

\subsection{The soft Higgsino mass}\label{sec:Higgsinomass}
Low fine-tuning requires that there should be no significant relation between uncorrelated coefficients of the soft terms. However in specific SUSY breaking schemes there may be correlations between the coefficients and such natural correlations can significantly affect the fine-tuning measure. For example if SUSY breaking leads to degenerate soft scalar masses there is a cancellation between the tree level and radiative contributions to the Higgs mass that leads to a reduction of the sensitivity of the Higgs mass to the initial scalar masses, the so-called ``focus point''. As a result the fine-tuning measure is significantly reduced.

Thus, when computing fine-tuning,  it is important to take care of all possible natural correlations between the coefficients of the soft terms. Here we argue that the soft Higgsino mass provides one such correlation that has not been included in fine tuning estimates and that it can lead to a significant reduction in fine-tuning.  The reason that it is not included is that it can be eliminated by a change in the supersymmetric ``$\mu$ term", $\mu H_uH_d|_{\theta\theta}$, together with a change in the Higgs soft masses\footnote{In general also the non-holomorphic trilinear couplings T' need to be shifted due to the $F$-term contribution of the superpotential $\mu$-term.}, $m_{H_u^2} |H_u|^2, m_{H_d^2}|H_d|^2$:
\be
\mu'\tilde{H_u}\tilde{H_d}\equiv m_{\tilde H}H_uH_d|_{\theta\theta}-m_{\tilde H}^2(|H_u|^2|+H_d|^2)
\ee
However dropping the Higgsino mass term is inconsistent with the determination of the fine-tuning measure because, as may be seen from this equation, the Higgsino mass term implies a natural correlation between the coefficients of the $\mu$ term and the Higgs soft masses.

Of course it is important to ask whether, in an effective field theory sense, an Higgsino mass can occur with a coefficient uncorrelated with the other soft SUSY breaking terms of the MSSM. It is straightforward to establish that this is the case. For example the authors of reference \cite{Antoniadis:2008es} have tabulated all the allowed dimension 5 operators in the MSSM that are  consistent with R parity.  In particular they find the operator
\bea
O&=&\frac{1}{M}\int d^4\theta [ A(S,S^\dag )D^\alpha \left( B(S,S^\dag )H_2 e^{ - V_1} \right)\nonumber\\
&&\times D_\alpha \left( \Gamma (S,S^\dag )e^{V_1}H_1\right) + h.c. ]
\eea
where $A,B$ and $\Gamma$ are functions of the SUSY breaking spurion $S=M_s\theta^2$ where $M_s$ is the SUSY breaking scale, $V_1$ is a combination of the MSSM vector superfields,  $M$ is the mediator mass coming from integrating out massive fields in the underlying theory. For example, this particular operator can be generated by integrating out two massive $SU(2)$ multiplets that are coupled to the MSSM Higgs supermultiplets and in this case $M$ is the mass of these massive doublets. Including the SUSY breaking effects this operator generates a soft Higgsino mass term with coefficient proportional to the coefficient of the $S S^\dagger$ term in $A$. As this is the only SUSY breaking term proportional to this coefficient the soft Higgsino mass is not correlated with other SUSY breaking terms, and should be included when calculating fine-tuning in the MSSM.

\section{The impact of the new  soft-terms on the fine-tuning measure}
\label{sec:FT}
The fine tuning measure which we consider with respect to a set of independent parameters, $p$,  is given by ~\cite{Ellis:1986yg, Barbieri:1987fn}
\begin{align} 
\label{eq:measure}
\Delta \equiv \max {\text{Abs}}\big[\Delta _{p}\big],\qquad \Delta _{p}\equiv \frac{\partial \ln
  v^{2}}{\partial \ln p} = \frac{p}{v^2}\frac{\partial v^2}{\partial p} \;.
\end{align}
The quantity $\Delta^{-1}$ gives a measure of the accuracy to which independent parameters must be tuned to get the correct electroweak breaking scale.
In the following we will concentrate on the contributions of $\mu$ and $\mu'$ on the fine tuning measure.

The generic expressions for the Renormalisation Group Equations  (RGEs) in the presence of non-holomorphic soft-terms are 
given in Refs.~\cite{Jack:1999ud,Jack:1999fa}. We have implemented them in the \Mathematica package \SARAH  \cite{Staub:2008uz,Staub:2009bi,Staub:2010jh,Staub:2012pb,Staub:2013tta,Staub:2015kfa} to calculate the $\beta$-functions for all relevant terms in the considered model. The one-loop results for running of the 
new holomorphic soft-terms as well as the standard soft-breaking masses are summarised in appendix~\ref{app:RGEs}. We can use these results to find 
an approximate dependence of the running $m_{H_u}^2$ as function of all other soft-breaking terms. For this purpose, we assume CMSSM-like 
boundary conditions at the scale $M_{GUT} = 2.0 \times 10^{16}$~GeV
\begin{align}
& M_1 = M_2 = M_3 \equiv m_{1/2} \,,\quad
 m_{H_d}^2 = m_{H_u}^2 \equiv m_0^2 \nonumber \\
& m_e^2=m_d^2=m_u^2=m_l^2=m_q^2 \equiv {\bf 1} m_0^2  \nonumber \\
& T_i = A_0 Y_i \,,\quad 
T'_i = A'_0 Y_i \nonumber
\end{align}
and expand around $m_0=m_{1/2}=A_0=A'_0=\mu=\mu'=1$~TeV.
For $\tan\beta=50$, we find
\begin{align}
& m^2_{H_u}(Q) \simeq 0.001 A_b A_t -0.002 A_b M_2-0.009 A_b M_3 \nonumber \\
& -0.007 {A'_b}^2 +0.002 {A'_b} {A'_t}-0.024 {A'_b} \mu'-0.013 {A'_\tau}^2 \nonumber \\
& -0.012 {A'_\tau} \mu'-0.032 A_t^2+0.007 A_t M_1 +0.039 A_t M_2 \nonumber \\
&+0.145 A_t M_3+0.015 {A'_t}^2+0.023 {A'_t} \mu'+0.007 M_1^2 \nonumber \\
& -0.005 M_1 M_2-0.021 M_1 M_3+0.222 M_2^2-0.131 M_2 M_3  \nonumber \\
&-1.479 M_3^2  -0.051 m^2_{d,12}-0.013 m^2_{d,3}-0.051 m^2_{e,12} \nonumber \\
&-0.026 m^2_{e,3}  +0.037 m_{H_d}^2+0.637 m_{H_u}^2+0.051 m^2_{l,12}\nonumber \\
&+0.025 m^2_{l,3}  -0.051 m^2_{q,12}-0.350 m^2_{q,3}+0.101 m^2_{u,12}\nonumber \\
& -0.287 m^2_{u,3}+0.168 \mu'^2
\end{align}   

Here, we neglected first and second generation Yukawa couplings and skipped all terms with coefficients smaller than $10^{-3}$. 
In addition, we parametrised $T^{(')}_i = A^{('})_i Y_i$ for $(i=t,b,\tau)$. For both $\mu$ and $\mu'$ we obtain the simple relation $\mu^{(')} \simeq 0.86 \mu^{(')}(M_\text{GUT})$.
Thus, the dependence of the Z-boson mass at the weak scale on the parameters at the GUT scale is given by
\begin{equation}
\frac{M^2_Z}{2} \simeq -0.17 \mu'(M_\text{GUT})^2 - 0.86 \mu(M_\text{GUT})^2 + \dots
\end{equation}
Neglecting mixing effects in the neutralino sector, the Higgsino mass is given by $m_{\tilde{H}} \sim 0.86 (\mu+\mu')$.
We therefore expect that the fine tuning can be very mild even for rather heavy Higgsinos, if the main contribution to its
mass comes from the non-holomorphic soft term.

\section{Parameter scan and plots}
 \label{sec:num}
We verified this expectation with an explicit numerical study. 
As free parameters at the GUT scale we take $m_0,m_{1/2},A_0, \tan\beta, \mu, B\mu, \mu'$. The correct electroweak vacuum is ensured via the choice of the
Higgs masses $m_{H_u}^2, m_{H_d}^2$. The overall fine tuning in this setup will typically not be dominated by $\mu$, but this could be changed by 
considering e.g.\ non-universal gaugino masses (see e.g.~\cite{Horton:2009ed,Kaminska:2013mya}). 
As these two problems `factorise' we concentrate on the fine tuning with respect to $\mu$ and $\mu'$.
 
\begin{figure}[!h!]
 \centering
 \includegraphics[width=0.9\linewidth]{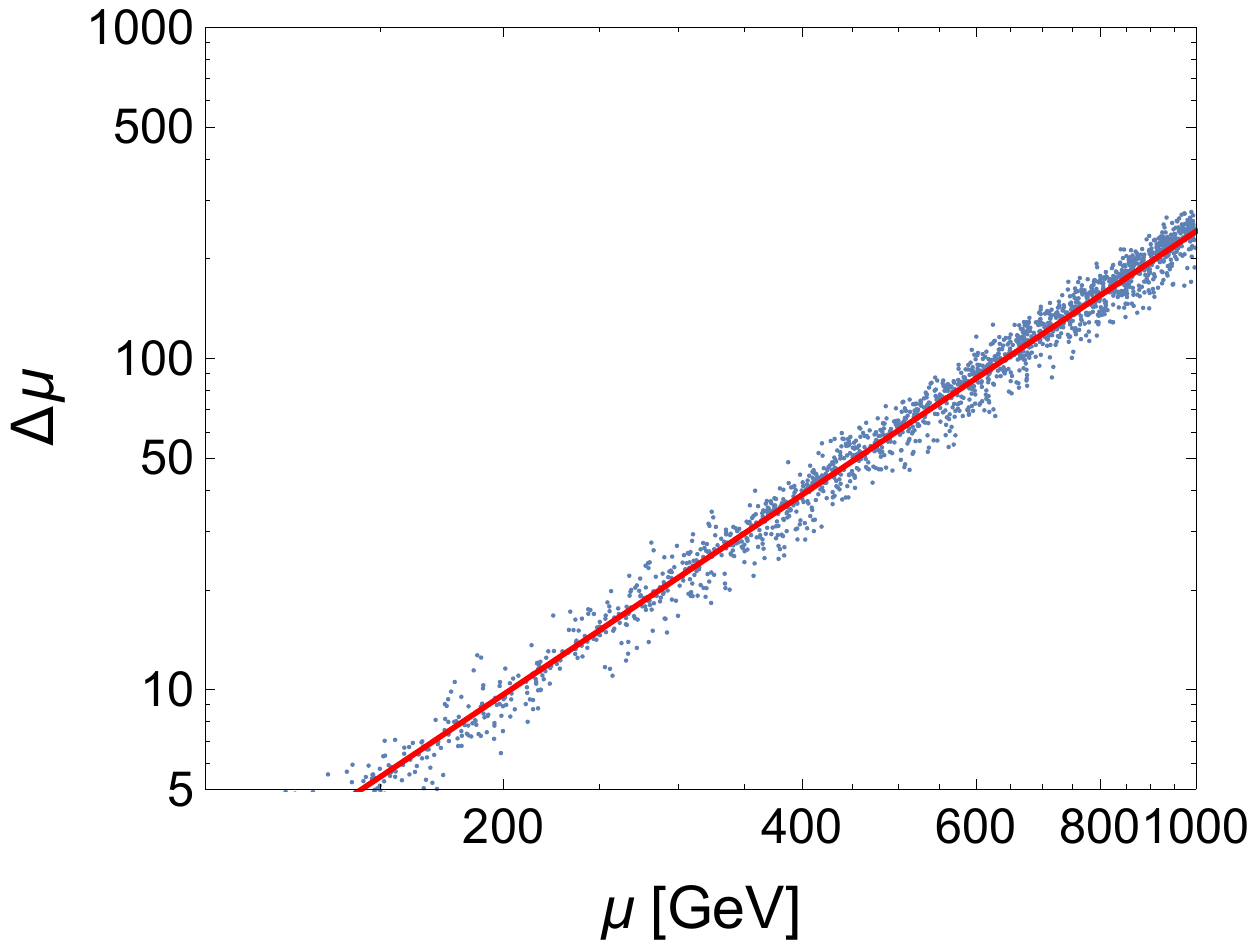}
    \includegraphics[width=0.9\linewidth]{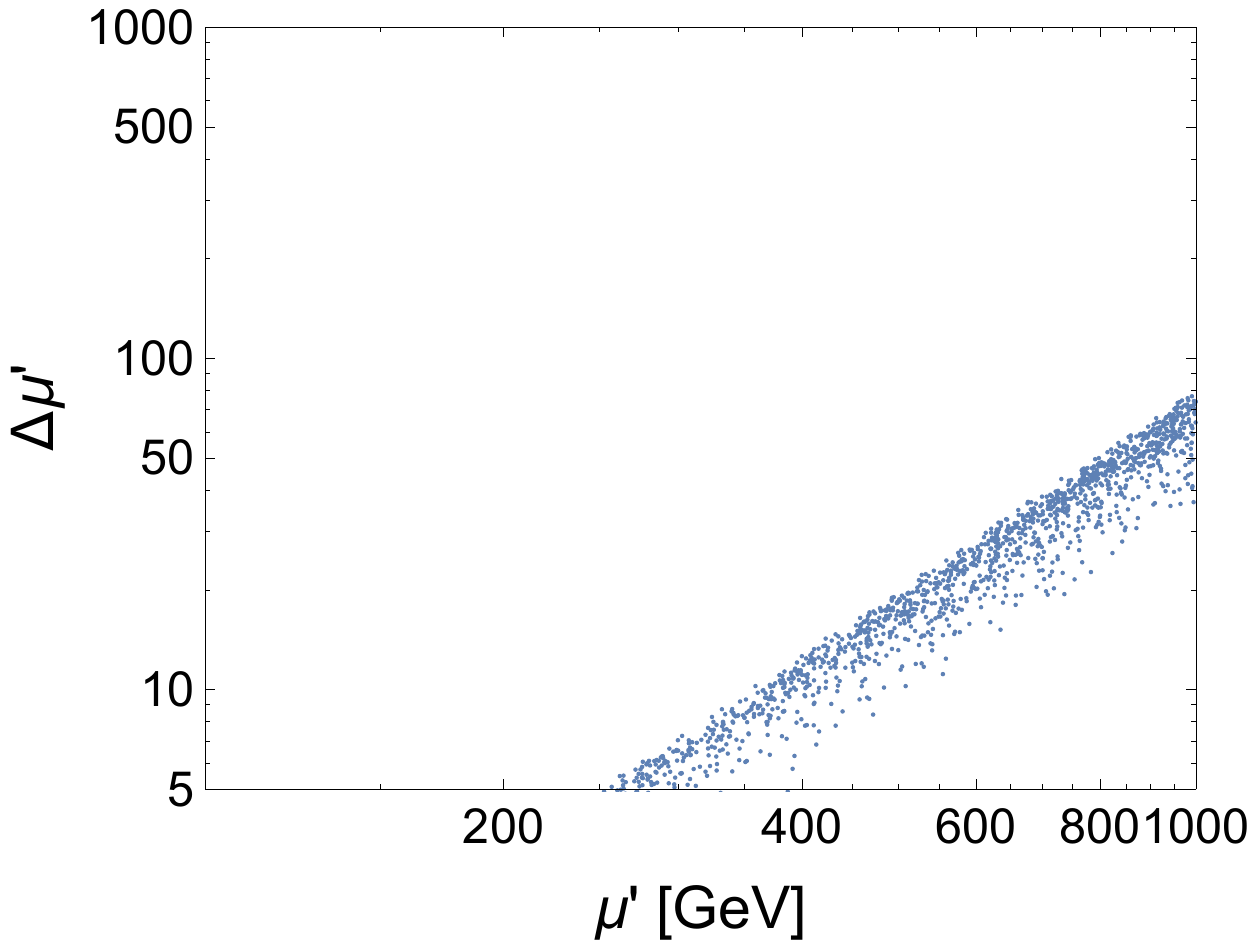} 
 \caption{Top: Contribution of $\mu$ to the fine tuning measure vs.\ the value of $\mu$. The red line
 corresponds to the rough estimate $\Delta_\mu \sim \frac{2\mu^2}{M_Z^2}$, which we observe to be an excellent approximation.
 Bottom: Contribution of $\mu'$ to the fine tuning measure vs.\ the value of $\mu'$. }
 \label{fig:FTmup}
 \end{figure}

\begin{figure}[!h!]
 \centering
 \includegraphics[width=0.9\linewidth]{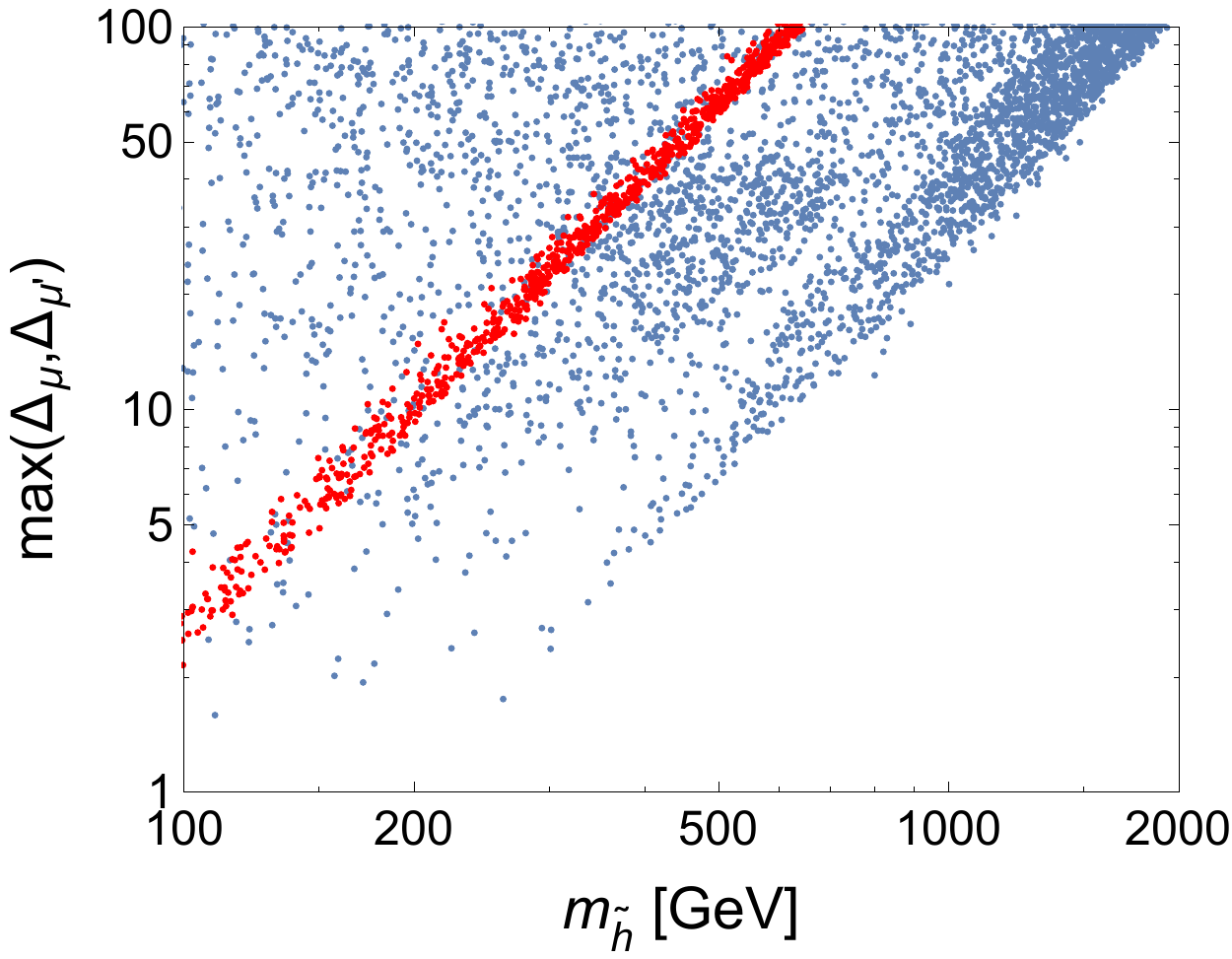}
 \caption{The maximum $\Delta_{\mu,\mu'}$ contribution to the fine tuning measure plotted against the Higgsino mass for varying $\mu'$ (blue) and $\mu'=0$ (red). }
 \label{fig:FThiggsino}
 \end{figure}

We performed a scan over the MSSM parameter space using the \SARAH generated \SPheno \cite{Porod:2011nf,Porod:2003um} version. In Fig.~\ref{fig:FTmup} we show the contribution to the fine tuning measure with respect to $\mu$  and the non-holomorphic
Higgsino mass term $\mu'$. We find that the usual approximation for the fine tuning with respect to $\mu$, $\Delta_\mu \sim \frac{2\mu^2}{M_Z^2}$
is an excellent approximation.
 Inspecting the plots we infer that the lowest fine tuning for a given Higgsino mass will be achieved for a non-holomorphic contribution to the Higgsino mass
 which is about 4-5 times larger than the contribution from the usual $\mu$ term. If we aim for a Higgsino mass of 1~TeV, particularly interesting from
 the dark matter perspective as it naturally gives the correct relic abundance, the ideal combination with respect to 
 fine tuning would therefore be for values $\mu\sim200$~GeV and $\mu'\sim 1000$~GeV, resulting in a fine tuning
 of about $\Delta_\mu\sim\Delta_{\mu'}\sim20$. Without the soft Higgsino mass term the fine tuning would be $\Delta_\mu \sim 350$!
This estimate is confirmed in Figure \ref{fig:FThiggsino} where we plot the mass of the neutralinos with the largest Higgsino component against the fine tuning, showing that indeed it can be of $\mathcal{O}(1~\Tev)$ for $\Delta_{\mu,\mu}'\le 20$. 
Depending on the parameter choice this state can be the LSP and give the correct relic abundance to be dark matter. For comparison we also show the fine tuning for the case where $\mu'=0$ (red points).

 \section{Summary and conclusions}
 \label{sec:conclusion}
 In this letter we have shown that a heavy Higgsino with a mass of $\mathcal{O}(1~\Tev)$ can arise in the MSSM without having a very large contribution to the the fine tuning in the MSSM from the $\mu$ term, provided that one allows for a non-holomorphic soft SUSY breaking Higgsino mass term. Such a heavy Higgsino has a sufficiently small annihilation cross section so that it can readily be dark matter without the need for any additional dark matter component. Although the soft Higgsino mass term is equivalent to a combination of a supersymmetric $\mu$ term and soft SUSY breaking Higgs mass terms, it is essential to keep the Higgsino mass term explicitly when calculating the fine tuning because it naturally correlates the magnitude of the equivalent  $\mu$ term and soft Higgs mass terms in such a way as to largely cancel the fine tuning contributions  of these terms.
 
 In order to make the role of the Higgsino mass clear,  we have concentrated on the contribution to fine tuning coming from the $\mu$ term and the soft Higgsino mass term. However the significant suppression of these contributions  that we find means that other contributions to the fine tuning are likely to be dominant and it will be important to perform a complete analysis including all fine tuning contributions.  In this context it will also be important to include the Higgsino soft mass term in extensions of the MSSM, such as those with non-universal gaugino masses \cite{Horton:2009ed,Kaminska:2013mya} or the generalised NMSSM \cite{Ross:2011xv,Ross:2012nr,Kaminska:2013mya}     that have been shown to reduce fine tuning. We hope to consider these issues shortly.
 
 \section*{acknowledgement}
This work is supported by the German Science Foundation (DFG) under the Collaborative 
Research Center (SFB) 676 Particles, Strings and the Early Universe as well as the ERC Starting Grant ‘NewAve’ (638528).
 
 \begin{appendix}
\section{Renormalisation group equations including non-holomorphic soft-terms}
\label{app:RGEs}
\subsection{RGEs for non-holomorphic soft-terms}
\begin{align}
\beta_{{T^{'}_u}}^{(1)} & =  
+3 {{T^{'}_u}  Y_{d}^{\dagger}  Y_d} +{{T^{'}_u}  Y_{u}^{\dagger}  Y_u}+2 {Y_u  Y_{d}^{\dagger}  {T^{'}_d}}  -4 {\mu'} {Y_u  Y_{d}^{\dagger}  Y_d}  \nonumber \\
&+2 {Y_u  Y_{u}^{\dagger}  {T^{'}_u}} -\frac{6}{5} Y_u \Big(\Big(5 g_{2}^{2}  + g_{1}^{2}\Big){\mu'}  -5 \mbox{Tr}\Big({{T^{'}_u}  Y_{u}^{\dagger}}\Big) \Big) \nonumber\\
 & +{T^{'}_u} \Big(3 \mbox{Tr}\Big({Y_d  Y_{d}^{\dagger}}\Big)  -\frac{4}{15} \Big(20 g_{3}^{2}  + g_{1}^{2}\Big) + \mbox{Tr}\Big({Y_e  Y_{e}^{\dagger}}\Big)\Big)\\ 
\beta_{{T^{'}_d}}^{(1)} & =  
+{{T^{'}_d}  Y_{d}^{\dagger}  Y_d}+3 {{T^{'}_d}  Y_{u}^{\dagger}  Y_u} +2 {Y_d  Y_{d}^{\dagger}  {T^{'}_d}} +2 {Y_d  Y_{u}^{\dagger}  {T^{'}_u}} \nonumber \\ 
 &-4 {\mu'} {Y_d  Y_{u}^{\dagger}  Y_u} +Y_d \Big(2 \mbox{Tr}\Big({{T^{'}_e}  Y_{e}^{\dagger}}\Big)  + 6 \mbox{Tr}\Big({{T^{'}_d}  Y_{d}^{\dagger}}\Big)  \nonumber \\ 
 &-\frac{6}{5} \Big(5 g_{2}^{2}  + g_{1}^{2}\Big){\mu'} \Big)+\frac{1}{15} {T^{'}_d} \Big(2 g_{1}^{2}  + 45 \mbox{Tr}\Big({Y_u  Y_{u}^{\dagger}}\Big)  -80 g_{3}^{2} \Big)\\ 
\beta_{{T^{'}_e}}^{(1)} & =  
+{{T^{'}_e}  Y_{e}^{\dagger}  Y_e}+2 {Y_e  Y_{e}^{\dagger}  {T^{'}_e}} +Y_e \Big(2 \mbox{Tr}\Big({{T^{'}_e}  Y_{e}^{\dagger}}\Big)  + 6 \mbox{Tr}\Big({{T^{'}_d}  Y_{d}^{\dagger}}\Big)  \nonumber \\ 
 &-\frac{6}{5} \Big(5 g_{2}^{2}  + g_{1}^{2}\Big){\mu'} \Big)+{T^{'}_e} \Big(3 \mbox{Tr}\Big({Y_u  Y_{u}^{\dagger}}\Big)  -\frac{6}{5} g_{1}^{2} \Big)\\ 
\beta_{{\mu'}}^{(1)} & =  
3 {\mu'} \mbox{Tr}\Big({Y_d  Y_{d}^{\dagger}}\Big)  -\frac{3}{5} {\mu'} \Big(5 g_{2}^{2}  -5 \mbox{Tr}\Big({Y_u  Y_{u}^{\dagger}}\Big)  + g_{1}^{2}\Big) \nonumber \\ 
 &+ {\mu'} \mbox{Tr}\Big({Y_e  Y_{e}^{\dagger}}\Big) 
\end{align}
\subsection{RGEs for soft-breaking masses}

\allowdisplaybreaks{
\begin{align} 
\beta_{m_q^2}^{(1)} & =  
-\frac{2}{15} g_{1}^{2} {\bf 1} |M_1|^2 -\frac{32}{3} g_{3}^{2} {\bf 1} |M_3|^2 -6 g_{2}^{2} {\bf 1} |M_2|^2 +2 m_{H_d}^2 {Y_{d}^{\dagger}  Y_d} \nonumber \\ 
 &+2 m_{H_u}^2 {Y_{u}^{\dagger}  Y_u} +2 {T_{d}^{\dagger}  T_d} +2 {T_{u}^{\dagger}  T_u} +2 {{T^{'}_{d}}^{T}  {T^{'}_d}^*} +2 {{T^{'}_{u}}^{T}  {T^{'}_u}^*} \nonumber \\ 
 &-4 |{\mu'}|^2 {Y_{d}^{T}  Y_d^*} -4 |{\mu'}|^2 {Y_{u}^{T}  Y_u^*} +{m_q^2  Y_{d}^{\dagger}  Y_d}+{m_q^2  Y_{u}^{\dagger}  Y_u}\nonumber \\ 
 &+2 {Y_{d}^{\dagger}  m_d^2  Y_d} +{Y_{d}^{\dagger}  Y_d  m_q^2}+2 {Y_{u}^{\dagger}  m_u^2  Y_u} +{Y_{u}^{\dagger}  Y_u  m_q^2}+\frac{ g_1 {\bf 1}\sigma}{3}\\ 
\beta_{m_l^2}^{(1)} & =  
-\frac{6}{5} g_{1}^{2} {\bf 1} |M_1|^2 -6 g_{2}^{2} {\bf 1} |M_2|^2 +2 m_{H_d}^2 {Y_{e}^{\dagger}  Y_e} +2 {T_{e}^{\dagger}  T_e} \nonumber \\ 
 &+2 {{T^{'}_{e}}^{T}  {T^{'}_e}^*} -4 |{\mu'}|^2 {Y_{e}^{T}  Y_e^*} \nonumber \\ 
 &+{m_l^2  Y_{e}^{\dagger}  Y_e}+2 {Y_{e}^{\dagger}  m_e^2  Y_e} +{Y_{e}^{\dagger}  Y_e  m_l^2}- g_1 {\bf 1}\sigma\\ 
\beta_{m_{H_d}^2}^{(1)} & =  
-\frac{6}{5} g_{1}^{2} {\mu'}^{2} -6 g_{2}^{2} {\mu'}^{2} -\frac{6}{5} g_{1}^{2} |M_1|^2 -6 g_{2}^{2} |M_2|^2 - g_1\sigma\nonumber \\ 
 &+6 \mbox{Tr}\Big({{T^{'}_u}  {T^{'}_{u}}^{\dagger}}\Big) +6 m_{H_d}^2 \mbox{Tr}\Big({Y_d  Y_{d}^{\dagger}}\Big) +2 m_{H_d}^2 \mbox{Tr}\Big({Y_e  Y_{e}^{\dagger}}\Big) \nonumber \\ 
 &+6 \mbox{Tr}\Big({T_d^*  T_{d}^{T}}\Big)  +2 \mbox{Tr}\Big({T_e^*  T_{e}^{T}}\Big) +6 \mbox{Tr}\Big({m_d^2  Y_d  Y_{d}^{\dagger}}\Big) \nonumber \\ 
 &+2 \mbox{Tr}\Big({m_e^2  Y_e  Y_{e}^{\dagger}}\Big) +2 \mbox{Tr}\Big({m_l^2  Y_{e}^{\dagger}  Y_e}\Big) +6 \mbox{Tr}\Big({m_q^2  Y_{d}^{\dagger}  Y_d}\Big) \\ 
\beta_{m_{H_u}^2}^{(1)} & =  
-\frac{6}{5} g_{1}^{2} {\mu'}^{2} -6 g_{2}^{2} {\mu'}^{2} -\frac{6}{5} g_{1}^{2} |M_1|^2 -6 g_{2}^{2} |M_2|^2 + g_1\sigma\nonumber \\ 
 &+6 \mbox{Tr}\Big({{T^{'}_d}  {T^{'}_{d}}^{\dagger}}\Big) +2 \mbox{Tr}\Big({{T^{'}_e}  {T^{'}_{e}}^{\dagger}}\Big) +6 m_{H_u}^2 \mbox{Tr}\Big({Y_u  Y_{u}^{\dagger}}\Big) \nonumber \\ 
 &+6 \mbox{Tr}\Big({T_u^*  T_{u}^{T}}\Big) +6 \mbox{Tr}\Big({m_q^2  Y_{u}^{\dagger}  Y_u}\Big) +6 \mbox{Tr}\Big({m_u^2  Y_u  Y_{u}^{\dagger}}\Big) \\ 
\beta_{m_d^2}^{(1)} & =  
-\frac{8}{15} g_{1}^{2} {\bf 1} |M_1|^2 -\frac{32}{3} g_{3}^{2} {\bf 1} |M_3|^2 +4 m_{H_d}^2 {Y_d  Y_{d}^{\dagger}} \nonumber \\ 
 &+4 {{T^{'}_d}^*  {T^{'}_{d}}^{T}} -8 |{\mu'}|^2 {Y_d^*  Y_{d}^{T}} +4 {T_d  T_{d}^{\dagger}} +2 {m_d^2  Y_d  Y_{d}^{\dagger}} \nonumber \\ 
 &+4 {Y_d  m_q^2  Y_{d}^{\dagger}} +2 {Y_d  Y_{d}^{\dagger}  m_d^2} +\frac{2 g_1 {\bf 1}\sigma}{3} \\ 
\beta_{m_u^2}^{(1)} & =  
-\frac{32}{15} g_{1}^{2} {\bf 1} |M_1|^2 -\frac{32}{3} g_{3}^{2} {\bf 1} |M_3|^2 +4 m_{H_u}^2 {Y_u  Y_{u}^{\dagger}}  \nonumber \\ 
 &+4 {{T^{'}_u}^*  {T^{'}_{u}}^{T}}-8 |{\mu'}|^2 {Y_u^*  Y_{u}^{T}} +4 {T_u  T_{u}^{\dagger}} +2 {m_u^2  Y_u  Y_{u}^{\dagger}}\nonumber \\ 
 & +4 {Y_u  m_q^2  Y_{u}^{\dagger}} +2 {Y_u  Y_{u}^{\dagger}  m_u^2} - \frac{4 g_1 {\bf 1}\sigma}{3} \\ 
\beta_{m_e^2}^{(1)} & =  
-\frac{24}{5} g_{1}^{2} {\bf 1} |M_1|^2 +2 \Big(2 m_{H_d}^2 {Y_e  Y_{e}^{\dagger}} +2 {{T^{'}_e}^*  {T^{'}_{e}}^{T}} \nonumber \\ 
 &-4 |{\mu'}|^2 {Y_e^*  Y_{e}^{T}}  +2 {T_e  T_{e}^{\dagger}} +{m_e^2  Y_e  Y_{e}^{\dagger}}+2 {Y_e  m_l^2  Y_{e}^{\dagger}} \nonumber \\ 
 &+{Y_e  Y_{e}^{\dagger}  m_e^2}\Big)+2 g_1 {\bf 1}\sigma
\end{align}  }
with 
\begin{align}
\sigma & = {\frac{3}{5}} g_1 \Big(-2 \mbox{Tr}\Big({m_u^2}\Big)  - \mbox{Tr}\Big({m_l^2}\Big)  - m_{H_d}^2  + m_{H_u}^2 \nonumber \\
& + \mbox{Tr}\Big({m_d^2}\Big) + \mbox{Tr}\Big({m_e^2}\Big) + \mbox{Tr}\Big({m_q^2}\Big)\Big) 
\end{align}

\end{appendix}

\bibliography{NMSSM}
\bibliographystyle{ArXiv}

\end{document}